\documentstyle[12pt]{article}

\newcommand{\mysection}{\setcounter{equation}{0}\section}

\begin{document}
\hfill{INLO-PUB-11/97}
\vskip 1.0cm
\centerline{\large\bf { O($\alpha_s^2$) Contributions to the fragmentation
function $g_1(x,Q^2)$ }}
\centerline{\large\bf {in polarized $e^+e^-$ -annihilation}}
\vskip 2.0cm
\centerline {\sc P.J. Rijken and W.L. van Neerven}
\vskip 1.0cm
\centerline{\it Instituut-Lorentz,}
\centerline{\it University of Leiden,}
\centerline{\it PO Box 9506, 2300 RA Leiden,}
\centerline{\it The Netherlands.}
\vskip 1.0cm
\centerline{November 1997}
\vskip 2.0cm
\centerline{\bf Abstract}
\vskip 0.3cm
We present the calculation of the order $\alpha_s^2$ contributions to the
coefficient functions (singlet and non-singlet) belonging to the longitudinal
spin fragmentation function $g_1(x,Q^2)$ measured in polarized 
electron-positron
annihilation. From this calculation we also obtain the two-loop
contribution to the timelike spin splitting functions $P_{qq}^{\rm S}$ and
$P_{gq}$. These splitting functions are in agreement with recent results
in the literature which were obtained by using a different method. We show
that in second order the renormalization constant needed for the HVBM 
prescription of the $\gamma_5$-matrix is process dependent.

\vfill
\newpage

\mysection{Introduction}
The measurement of the fragmentation functions $F_i^{\rm H}(x,Q^2)$ in 
unpolarized
electron-positron annihilation provides us with a new possibility to test
the predictions of perturbative quantum chromodynamics (QCD). Here the
reaction under consideration is given by $e^+ + e^- \rightarrow {\rm H} + "X"$
where H represents the detected hadron and $"X"$ stands for any inclusive
hadronic state. Because of the similarity between the above process and 
deep inelastic lepton-hadron scattering $e^- + {\rm H} \rightarrow e^- + "X"$
one can consider the fragmentation functions as the timelike analogues of
the well known deep inelastic structure functions measured in the latter 
process. This similarity becomes clear if one looks at the experimental
and the theoretical characteristics of both functions. As far as experiment
is concerned one has a full control of the background and a large amount of
data so that the systematical and statistical errors are rather small. The
data are collected over a wide range of $Q^2$ values which enable us to
test the scaling violations exhibited by both functions (for the fragmentation
function see e.g. \cite{abreu}, \cite{busk1}, \cite{akers}. 
From the theoretical viewpoint
structure functions and fragmentation functions have the common feature that
the Born approximation is determined by the electroweak interactions only.
Therefore it is easier to calculate their QCD corrections at higher order in  
the strong coupling constant $\alpha_s$ than for other quantitities.
It turns out that the most complicated virtual corrections to these reactions, 
which are represented by  
n- point functions with $n \ge 4$, appear in much higher order than 
for other processes like e.g. jet production.
Furthermore the semi inclusive nature of the above processes leads to a
considerable simplification of the phase space integrals. Another important 
feature is that one does not have to deal
with problems like jet definitions and hadronization effects in the case of 
inclusive quantities.
Because of the these properties it was possible to compute the order 
$\alpha_s^2$
contributions to the coefficient functions corresponding to the structure
functions \cite{zn1}, \cite{zn2}, \cite{zn3} and the fragmentation functions 
\cite{rn1}, \cite{rn2}, \cite{rn3},
\cite{rijk}. In spite of the similarities between the semi inclusive
quantities in deep inelastic scattering and eletron-positron annihilation there
is also one striking difference. This can be attributed to the current current
correlation function which shows up in the expressions for the structure
functions as well as in the fragmentation functions. In the case of deep 
inelastic 
scattering the current current correlation function is dominated by the 
light cone in the Bj{\o}rken limit whereas in the case of electron-positron 
annihilation this property does not exist. Therefore the light cone expansion
techniques cannot be applied to the fragmentation functions. These techniques
allow us to relate parton densities to the expectation values of hadronic
operators between hadronic states. It also enables us to compute sum rules
of structure functions to even higher orders in perturbation theory than is
possible for the coefficient functions (see e.g. \cite{lave}).

Up to now we have only discussed unpolarized scattering. It would be 
interesting to see whether QCD also yields good predictions when the hadron
H is polarized. This has been investigated theoretically as well as
experimentally during the last ten years for polarized deep inelastic
scattering. The most effort was devoted to the longitudinal spin structure 
function
$g_1(x,Q^2)$ because of the disagreement between the data \cite{ash} and its 
first moment given by the Ellis-Jaffe sum rule \cite{elja}.
This discrepancy lead to a thorough investigation from which one can conclude
that it is more difficult to make QCD predictions for spin dependent quantities 
than for spin averaged ones. Therefore
we expect that the same will happen for electron-positron annihilation
when the detected hadron ${\rm H}$ in the final state is polarized. This 
process allows us to study the timelike analogue of the above longitudinal spin
structure function which we will denote by  the fragmentation function 
$g_1^{\rm H}(x,Q^2)$. 
Contrary to polarized deep inelastic scattering there do not exist 
data for the longitudinal spin fragmentation function. At this moment there
are some data on polarized $\Lambda$-baryon production (see e.g. \cite{busk2}) 
so that in the future one maybe able to measure 
the fragmentation densities $\Delta D_a^{\Lambda}(z,\mu^2)$ for $a=q,\bar q,g$
which appear in $g_1^{\Lambda}(x,Q^2)$.
On the theoretical side considerable activity has been devoted in the past few
years to understand the fragmentation in polarized $\Lambda$-baryon production  
in spacelike as well as in timelike processes. For recent work we refer to 
\cite{buja}, \cite{ravin1}, \cite{ravin2}, \cite{flsa}, \cite{stvo} 
(for older work see also \cite{raba}). Apart from the 
nonperturbative aspects it is also important to calculate the QCD corrections
to the spin fragmentation functions. Recent progress has been made in the
computation of the next-to-leading order (NLO) corrections to $g_1^H(x,Q^2)$.
The order $\alpha_s$ contributions to the coefficient functions have been 
calculated in \cite{ravin1}-\cite{stvo} and the two-loop contributions to the 
timelike spin splitting functions are presented for the first time in 
\cite{stvo}. It is our aim to extend the above calculations by including 
the order $\alpha_s^2$ contributions to the coefficient functions which is
a first step to obtain the next-to-next-to-leading (NNLO) corrections to 
$g_1^{\rm H}(x,Q^2)$. 

Our paper will be organized as follows. In the next 
section we introduce our notations and present the framework of the computations
of the longitudinal spin timelike coefficient functions. In section 3 we
give an outline of the calculation. From the latter we also obtain two out of
the four splitting functions computed in \cite{stvo}. Further we discuss
the intricacies of the prescription of the $\gamma_5$-matrix in n dimensions 
and show the process dependence
of the renormalization constant needed to restore the 
Ward-identities when this matrix is treated in the scheme of HVBM 
\cite{hv}, \cite{bm}.
The long expressions for the order $\alpha_s^2$ corrected coefficient 
functions are presented in appendix A. 


\newcommand{\ps}{p\hspace{-0.42em}/\hspace{0.1em}}
\mysection{Fragmentation functions in polarized and unpolarized
electron-positron annihilation}
Single particle inclusive production in electron-positron
annihilation is given by the reaction
\begin{eqnarray}
\label{eqn:2.1}
e^-(\lambda_1,l_1) + e^+(\lambda_2,l_2) \rightarrow V(q) \rightarrow 
{\rm H}(s,p) + "{\rm X}" \,,
\end{eqnarray}
where $"{\rm X}"$ represents any inclusive final hadronic state.
Further $\lambda_i,l_i$ ($i=1,2$) denote the spin and momentum of the
incoming leptons and $s,p$ stand for the spin and momentum of the hadron H
detected in the final state. If we denote the momentum of the virtual vector
boson $V$, with $V=\gamma$ or $V=Z$, by $q=l_1+l_2$ the Bj{\o}rken scaling 
variable is defined by
\begin{eqnarray}
\label{eqn:2.2}
x= \frac{2p\cdot q}{Q^2}\,, \quad q^2=Q^2 > 0\,, \quad 0 < x \leq 1 \,.
\end{eqnarray}
The spin averaged differential cross section can be written as
\begin{eqnarray}
\label{eqn:2.3}
\frac{d\sigma^{\rm H}}{dxd\Omega}=\frac{N}{4}\sum_V \sum_{V'} {\cal L}_{(V,V')}^
{\mu\nu}(l_1;l_2) {\cal D}_{(V,V')}(Q^2) 
W_{\mu\nu}^{average,{\rm H},(V,V')}(p,q) \,,
\end{eqnarray}
where $V$ and $V'$ run over $\gamma$ and $Z$ and $N$ stands for the colour 
factor of the local gauge group $SU(N)$. Further we have denoted the
spin averaged leptonic tensor by ${\cal L}_{(V,V')}^{\mu\nu}(l_1;l_2)$ and
${\cal D}_{(V,V')}(Q^2)$ stands for the electroweak vector boson propagators.
If we average over the spin of the detected hadron H one obtains the
hadronic tensor $W_{\mu\nu}^{average,{\rm H},(V,V')}(p,q)$ which can be 
expressed
into the fragmentation functions $F_i^{{\rm H},(V,V')}(x,Q^2)$ (see below). 
An explicit
expression for Eq. (\ref{eqn:2.3}) can be found in \cite{rn2} and \cite{nw}.
In the case the electron is polarized we can choose $\lambda_1=\downarrow$ and
average over the spin $\lambda_2$ of the positron in Eq. (\ref{eqn:2.1}). The
difference between the cross sections where the detected hadron is polarized
parallel $s=\downarrow$ and antiparallel $s=\uparrow$ with respect to the 
electron is given by
\begin{eqnarray}
\label{eqn:2.4}
&&\frac{d\sigma^{{\rm H}(\downarrow)}(\downarrow)}{dxd\Omega}-
\frac{d\sigma^{{\rm H}(\uparrow)}(\downarrow)}{dxd\Omega}=
\nonumber\\
&& \frac{N}{2}\sum_V \sum_{V'} {\cal L}_{(V,V')}^
{\mu\nu}(\downarrow,l_1;l_2) {\cal D}_{(V,V')}(Q^2) 
W_{\mu\nu}^{spin,{\rm H},(V,V')}(s,p,q) \,.
\end{eqnarray}
For a polarized electron the leptonic tensor is given by
${\cal L}_{(V,V')}^{\mu\nu}(\downarrow,l_1;l_2)$ and the spin dependent hadronic
tensor $W_{\mu\nu}^{spin,{\rm H},(V,V')}(s,p;q)$ can be expressed into the spin
fragmentation functions $g_i^{{\rm H},(V,V')}(x,Q^2)$ (see below). Explicit 
formulae for Eq. (\ref{eqn:2.4}) can be found in \cite{ravin2}. In general the
hadronic tensor can be written as
\begin{eqnarray}
\label{eqn:2.5}
&&W_{\mu\nu}^{{\rm H},(V,V')}(s,p,q)=
\nonumber\\
&& \frac{1}{4\pi} \sum_X \int \,d^4y \, e^{iqy} \, \langle 0 \mid 
J_{\mu}^V(y) \mid {\rm H}(s,p), X \rangle 
\langle {\rm H}(s,p), X \mid J_{\nu}^{V'}(0)
\mid 0\rangle
\nonumber\\
&&=W_{\mu\nu}^{average,{\rm H},(V,V')}(p,q)
+W_{\mu\nu}^{spin,{\rm H},(V,V')}(s,p,q)\,.
\end{eqnarray}
The spin average hadronic tensor can be expressed into the fragmentation 
functions $F_i(x,Q^2)$ as follows
\begin{eqnarray}
\label{eqn:2.6}
&& W_{\mu\nu}^{average,{\rm H},(V,V')}(p,q) =
\Big (-g_{\mu\nu}+\frac{q_{\mu}q_{\nu}}{q^2}
\Big ) F_1^{{\rm H},(V,V')}(x,Q^2) 
\nonumber\\
&& + \frac{\hat {p}_{\mu}\hat {p}_{\nu}}{p\cdot q}
F_2^{{\rm H},(V,V')}(x,Q^2)
 + i \varepsilon_{\mu\nu\lambda\sigma}\frac{q^{\lambda}p^{\sigma}}{2p\cdot q}
F_3^{{\rm H},(V,V')}(x,Q^2) \,.
\end{eqnarray}
Similarly one can write the spin dependent hadronic tensor as
\begin{eqnarray}
\label{eqn:2.7}
&& W_{\mu\nu}^{spin,{\rm H},(V,V')}(s,p,q)=
\nonumber\\
&& i \varepsilon_{\mu\nu\lambda\sigma}\frac{q^{\lambda}s^{\sigma}}{2p\cdot q}
g_1^{{\rm H},(V,V')}(x,Q^2) + 
i \varepsilon_{\mu\nu\lambda\sigma}q_{\lambda}\Big (s_{\sigma}-
\frac{s\cdot q}{p\cdot q} p_{\sigma}\Big )\frac{g_2^{{\rm H},(V,V')}(x,Q^2)}
{2p\cdot q}
\nonumber\\
&&+  \Big (\frac{\hat {p}_{\mu}\hat {s}_{\nu}+\hat {s}_{\mu}\hat {p}_{\nu}}{2}-
s\cdot q\frac{\hat {p}_{\mu}\hat {p}_{\nu}}{p\cdot q}\Big )
\frac{g_3^{{\rm H},(V,V')}(x,Q^2)}{p\cdot q}
+  s\cdot q \frac{\hat {p}_{\mu}\hat {p}_{\nu}}{(p\cdot q)^2} 
g_4^{{\rm H},(V,V')}(x,Q^2) 
\nonumber\\
&&+ \Big (-g_{\mu\nu}+\frac{q_{\mu}q_{\nu}}{q^2}\Big )\frac{s\cdot q}{p\cdot q}
g_5^{{\rm H},(V,V')}(x,Q^2) \,,
\end{eqnarray}
where $g_i^{\rm H}(x,Q^2)$ denote the spin fragmenation functions. 
Further we have 
introduced the abreviations
\begin{eqnarray}
\label{eqn:2.8}
\hat {p}_{\mu}=p_{\mu}-\frac{p\cdot q}{q^2} q_{\mu} \quad , \quad
\hat {s}_{\mu}=s_{\mu}-\frac{s\cdot q}{q^2} q_{\mu} \,.
\end{eqnarray}
The above fragmentation functions can be written as a product of the 
electroweak couplings, describing the interaction of the vector boson $V$ with 
the quarks, and the fragmentation functions $F_i^{\rm H}$, $g_i^{\rm H}$ so   
that the superscripts $(V,V')$ can be dropped in the latter quantities.
The QCD analysis of the fragmentation functions proceeds in the same way as is
done for the structure functions in deep inelastic lepton-hadron scattering.
Hence it follows that the leading contributions to the fragmentation
functions $F_i^{\rm H}$ are all of type twist two. The same holds
for the spin fragmentation functions $g_i^{\rm H}$ ($i=1,4,5)$. 
However $g_2^{\rm H}$
and $g_3^{\rm H}$ also receive leading contributions of type twist three. This
means that the latter cannot be simply expressed into convolutions of
coefficient functions with fragmentation densities as can be done for
twist two contributions. Limiting ourselves to
twist two one can simply put $s=p$ in Eq. (\ref{eqn:2.7}) so that the functions
$g_2^{\rm H}$ and $g_3^{\rm H}$ drop out of the equations and we are left with  
the longitudinal spin quantities $g_i^{\rm H}$ with ($i=1,4,5$). Further 
one can derive 
\begin{eqnarray}
\label{eqn:2.9}
F_3^{\rm H}(x,Q^2)=-F_3^{\bar {\rm H}}(x,Q^2) \quad , \quad 
g_i^{\rm H}(x,Q^2)=-g_i^{\bar {\rm H}}(x,Q^2) \quad , \quad
\quad \hbox{$i=4,5$} \,,
\end{eqnarray}
from charge conjugation invariance of the strong interactions.
Hence it follows that the above fragmentation functions only receive
non-singlet (with respect to flavour) contributions. The other quantities,
given by $F_1^{\rm H}$, $F_2^{\rm H}$ and $g_1^{\rm H}$, also have singlet 
parts.
The QCD corrections to the spin averaged fragmentation functions $F_i^{\rm H}$
($i=1,2,3$) have been calculated up to second order in the strong coupling
constant $\alpha_s$ in \cite{rn1}, \cite{rn2}, \cite{rn3}, \cite{rijk} (for the
longitudinal fragmentation function $F_L^{\rm H}=F_2^{\rm H}-2xF_1^{\rm H}$ 
see also \cite{bin}).
Up to order $\alpha_s^2$ the QCD corrections to $g_4^{\rm H}$ and 
$g_5^{\rm H}$ are the
same as for the non-singlet parts of $F_2^{\rm H}$ and $F_1^{\rm H}$ 
respectively. The same
holds for the nonsinglet part of $g_1^{\rm H}$ which is equal to the 
correction to $F_3^{\rm H}$.
These equalities will be explained below. What is left is the singlet
contribution to $g_1^{\rm H}$ which will be calculated up to order 
$\alpha_s^2$ in 
this paper. This fragmentation function is the only one which shows up
in the spin dependent differential cross section (\ref{eqn:2.4}) if $V=\gamma$
so that process (\ref{eqn:2.1}) is purely electromagnetic.
In this case Eq. (\ref{eqn:2.4}) simplifies enormously and 
one gets
\begin{eqnarray}
\label{eqn:2.10}
\frac{d\sigma^{{\rm H}(\downarrow)}(\downarrow)}{dx\,d\cos\theta}-
\frac{d\sigma^{{\rm H}(\uparrow)}(\downarrow)}{dx\,d\cos\theta}=
N\frac{\pi \alpha^2}{Q^2} \,\cos\, \theta \, g_1^{\rm H}(x,Q^2) \,.
\end{eqnarray}
after integration over the azimuthal angle.
Notice that $g_1(x,Q^2)$ can be also measured in the reaction where both the
electron and positron are unpolarized. In this case it appears in the
cross section (\ref{eqn:2.3}) because of the axial vector coupling of the
Z-boson to the incoming leptons. In the purely electromagnetic case the
longitudinal spin fragmentation function can be written as 
\begin{eqnarray}
\label{eqn:2.11}
&& g_1^{\rm H}(x,Q^2)=\frac{1}{n_f}\sum_{k=1}^{n_f} e_k^2 \int_x^1 
\frac{dz}{z}\Big [
\Delta D_q^{\rm H,S}\Big (\frac{x}{z},M^2 \Big ) \Delta {\cal C}_{1,q}^{\rm S}
\Big (z,\frac{Q^2}{M^2} \Big )
\nonumber\\
&& + \Delta D_g^{\rm H,S}\Big (\frac{x}{z},M^2 \Big ) 
\Delta {\cal C}_{1,g}^{\rm S} \Big (z,\frac{Q^2}{M^2} \Big)
\nonumber\\
&& + n_f \Delta D_k^{\rm H,NS}\Big (\frac{x}{z},M^2 \Big ) 
\Delta {\cal C}_{1,q}^{\rm NS}\Big (z,\frac{Q^2}{M^2} \Big )
\Big ] \,.
\end{eqnarray}
The spin parton fragmentation densities are defined by
\begin{eqnarray}
\label{eqn:2.12}
\Delta D_a^{\rm H}(z,M^2)= D_{a\downarrow}^{\rm H\downarrow}(z,M^2) - 
D_{a\downarrow}^{\rm H\uparrow}(z,M^2) \,,
\end{eqnarray}
with $a=q,\bar q,g$. The quark fragmentation densities can be distinguished
into singlet (S) and non-singlet (NS) parts. The former is given by
\begin{eqnarray}
\label{eqn:2.13}
\Delta D_q^{\rm H,S}(z,M^2)=\sum_{k=1}^{n_f} \Big ( 
\Delta D_k^{\rm H}(z,M^2)+
\Delta D_{\bar k}^{\rm H}(z,M^2)\Big ) \,.
\end{eqnarray}
Here $k=1,2,\cdots n_f$ means $k=u,d,\cdots n_f$ where $n_f$ stands for the
heaviest light flavour appearing in the sums in Eqs. (\ref{eqn:2.11}),
(\ref{eqn:2.13}).
The spin non-singlet quark fragmentation densities are denoted by
\begin{eqnarray}
\label{eqn:2.14}
\Delta D_k^{\rm H,NS}(z,M^2)= \Delta D_k^{\rm H}(z,M^2)+\Delta 
D_{\bar k}^{\rm H}(z,M^2) - \frac{1}{n_f}
\Delta D_q^{\rm H,S}(z,M^2) \,.
\end{eqnarray}
The fragmentation densities and the coefficient functions 
$\Delta {\cal C}_{1,a}(z,Q^2/M^2)$ ($a=q,g$) depend 
on the partonic Bj{\o}rken scaling
variable $z$ (see below) and on the factorization scale
$M$ which we have put equal to the renormalization scale.

The coefficient functions are computed from the partonic subprocesses which
can be generally denoted by the reaction
\begin{eqnarray}
\label{eqn:2.15}
V(q) \rightarrow "a(s,p)" + a_1(s_1,p_1) + a_2(s_2,p_2) + \cdots + a_k(s_k,p_k)
\,,
\end{eqnarray}
where $a$ represents the detected parton ($a=q,\bar q,g$) with momentum $p$ 
which fragments into the hadron H.
The remaining partons $a_k$ belong to the inclusive state and one has to 
integrate over all momenta $p_k$ and to sum over all spins $s_k$. 
The reaction in Eq. (\ref{eqn:2.15})
can be described by a partonic structure tensor, indicated by a hat,
 which has the same form
as Eq. (\ref{eqn:2.5}) where ${\rm H}(s,p)$ and $X$ are replaced by $a(s,p)$ and
$\{a_k\}$ respectively. 
In the case the detected parton in Eq. (\ref{eqn:2.15}) is given
by a massless (anti-) quark the partonic structure tensor for unpolarized
scattering can be written as
\begin{eqnarray}
\label{eqn:2.16}
\hat W_{\mu\nu}^{average,q,(V,V')}(p,q)= {\rm Tr} \ps G_{\mu\nu}(p,q)\,.
\end{eqnarray}
In the above equation $G_{\mu\nu}$ denotes the amplitude squared where the
Dirac spinors of the detected quark $q(p)$ are removed. The indices $\mu$ and
$\nu$ refer to the (axial-) vector couplings of the vector bosons $V$ and 
$V'$ respectively. Further we have integrated over all momenta and summed over
the spins of the inclusive state in reaction (\ref{eqn:2.15}). 
If the gluon is detected in the final state of reaction (\ref{eqn:2.15}) we
can write 
\begin{eqnarray}
\label{eqn:2.17}
\hat W_{\mu\nu}^{average,g,(V,V')}(p,q)=P_{\alpha\beta}
G_{\mu\nu}^{\alpha\beta}(p,q) \,.
\end{eqnarray}
Here $G_{\mu\nu}^{\alpha\beta}$ is defined in an analogous way as 
$G_{\mu\nu}$ above and $\alpha,\beta$ refer to the Lorentz indices
of the polarization
vectors of the detected gluon i.e. $\epsilon^{\alpha}(p)$, $\epsilon^{\beta}(p)$
which are removed from $G_{\mu\nu}^{\alpha\beta}$. 
In unpolarized scattering $P_{\alpha\beta}$ stands for the spin averaged sum
over the polarizations. The same can be done for polarized scattering.
Choosing $s=p$ the partonic spin structure tensor reads
\begin{eqnarray}
\label{eqn:2.18}
\hat W_{\mu\nu}^{spin,q,(V,V')}(s,p,q)\mid_{s=p}= 
{\rm Tr}\gamma_5 \ps G_{\mu\nu}(p,q) \,.
\end{eqnarray}
If the gluon is detected in the final state of reaction (\ref{eqn:2.15}) we
can write 
\begin{eqnarray}
\label{eqn:2.19}
\hat W_{\mu\nu}^{average,g,(V,V')}(p,q)=\Delta P_{\alpha\beta} 
G_{\mu\nu}^{\alpha\beta}(p,q)\,.
\end{eqnarray}
In polarized scattering $\Delta P_{\alpha\beta}$ stands for 
the difference between the right- and the lefthanded 
polarizations. Here we will choose (see \cite{zn3})
\begin{eqnarray}
\label{eqn:2.20}
\Delta P_{\alpha\beta}=\frac{1}{p\cdot q}
\varepsilon_{\alpha\beta\sigma\lambda}p^{\sigma}q^{\lambda} \,.
\end{eqnarray}
From Eqs. (\ref{eqn:2.16})-(\ref{eqn:2.19}) one can infer the partonic 
fragmentation functions $\hat {\cal F}_{i,a}(z,Q^2)$ ($i=1,2,3$) and
$\hat g_{i,a}(z,Q^2)$ ($i=1,4,5$) with $a=q,g$ and $z=2p\cdot q/Q^2$
(see Eq. (\ref{eqn:2.15})) is the partonic Bj{\o}rken scaling variable. They 
are defined in a similar way as the hadronic fragmentation functions in
Eqs. (\ref{eqn:2.6}), (\ref{eqn:2.7}). Notice that the partonic fragmentation 
functions are collinearly divergent due to the higher order QCD radiative 
corrections. These divergences have to be removed via mass factorization so 
that one obtains the coefficient functions as e.g. shown in Eq. 
(\ref{eqn:2.11}). The calculation of the spin averaged quantities 
$\hat {\cal F}_{i,a}$ has been carried out up to order $\alpha_s^2$ in 
\cite{rn1}, \cite{rn2}, \cite{rn3}, \cite{rijk}. Here one has applied the 
method 
of n-dimensional regularization to compute the many body phase space integrals
and the loop integrals appearing in $G_{\mu\nu}$ (Eq. (\ref{eqn:2.16})) and
$G_{\mu\nu}^{\alpha\beta}$ (Eq. (\ref{eqn:2.17})). We will use the same method
to compute the partonic spin fragmentation function $\hat g_{1,a}$ 
which will be presented in the next section. In this case one has to find a
suitable prescription for the $\gamma_5$-matrix and the Levi-Civita tensor
in Eq. (\ref{eqn:2.20}). Here we have adopted the HVBM prescription given by
't Hooft and Veltman \cite{hv} which has been worked out in more detail by
Breitenlohner and Maison \cite{bm} (see also \cite{larin}, \cite{akde}).
The computation of $\hat g_{1,a}$ is completely analogous to its
deep inelastic counterpart described in \cite{zn3} and the spin averaged
partonic fragmentation functions $\hat {\cal F}_{1,a}$ in \cite{zn3} so that 
we do not have to
repeat the details in this paper. Before presenting the results in the next
section one can derive several relations between the coefficient functions
in the polarized and unpolarized case. However they are only valid
up to order $\alpha_s^2$. The non-singlet coefficient functions satisfy the
following equalities
\begin{eqnarray}
\label{eqn:2.21}
\Delta {\cal C}_{1,q}^{\rm NS}={\cal C}_{1,q}^{\rm NS} \quad , \quad 
\Delta {\cal C}_{4,q}^{\rm NS}={\cal C}_{2,q}^{\rm NS} \quad , \quad
\Delta {\cal C}_{5,q}^{\rm NS}={\cal C}_{1,q}^{\rm NS} \,,
\end{eqnarray}
For the singlet parts we have
\begin{eqnarray}
\label{eqn:2.22}
\Delta {\cal C}_{4,q}^{\rm PS}=0 \quad , \quad \Delta {\cal C}_{5,q}^{\rm PS}=0
\nonumber\\[2ex]
\Delta {\cal C}_{4,g}^{\rm S}=0 \quad ,\quad \Delta {\cal C}_{5,g}^{\rm S}=0 \,,
\end{eqnarray}
with the definition
\begin{eqnarray}
\label{eqn:2.23}
\Delta {\cal C}_{i,q}^{\rm S}=\Delta {\cal C}_{i,q}^{\rm NS}
+ \Delta {\cal C}_{i,q}^{\rm PS} \,.
\end{eqnarray}
The same relations also hold for the partonic fragmentation functions from
which the coefficient functions  are derived via mass factorization. The
relations in Eq. (\ref{eqn:2.21}) follow from the anticommutativity of the
$\gamma_5$-matrix which is manifest for a regularization method in
four dimensions. In the case of n-dimensional regularization it occurs after
the introduction of additional renormalization constants which restore
the various Ward-identities violated by the prescription used for the 
$\gamma_5$-matrix and the Levi-Civita tensor. The partonic fragmentation 
function $\hat g_{1,q}^{\rm NS}$ is derived from Eq. (\ref{eqn:2.18}) where
the indices $\mu$,$\nu$ either stand for $\gamma_{\mu}$,$\gamma_{\nu}$ or
for $\gamma_{\mu}\gamma_5$, $\gamma_{\nu}\gamma_5$. Further if $G_{\mu\nu}$
originates from so called rainbow graphs (see e.g. fig. 1a in \cite{lave})
then the $\gamma_5$-matrix in Eq. (\ref{eqn:2.18}) can be anticommuted in the
string of $\gamma$-matrices so that either the index $\mu$ or the index $\nu$
represents an axial-vector. In this way Eq. (\ref{eqn:2.18}) becomes equal
to Eq. (\ref{eqn:2.16}) which leads to ${\cal F}_{3,q}^{\rm NS}$. In the
case that either $\mu$ or $\nu$ (but not both) represent $\gamma_{\mu}\gamma_5$
or $\gamma_{\nu}\gamma_5$ one can anticommute the $\gamma_5$-matrix in
Eq. (\ref{eqn:2.18}) till it annihilates the other one in $G_{\mu\nu}$ and one
obtains Eq. (\ref{eqn:2.16}). In this way one obtains $\hat g_{4,q}^{\rm NS}=
\hat {\cal F}_{2,q}^{\rm NS}$ and $\hat g_{5,q}^{\rm NS}= 
\hat {\cal F}_{1,q}^{\rm NS}$. The relations in Eq. (\ref{eqn:2.22}) follow
from charge conjugation invariance of QCD so that one gets
\begin{eqnarray}
\label{eqn:2.24}
\hat {\cal F}_{3,a}= - \hat {\cal F}_{3,\bar a} \quad , \quad
\hat g_{i,a}= - \hat g_{i,\bar a} \quad , \quad  \mbox{$i=4,5$} \,.
\end{eqnarray}
Notice that the relations in Eq. (\ref{eqn:2.21}) do not hold anymore
beyond order $\alpha_s^2$. A nice example is given by the graphs in  fig. 1b of
\cite{lave} (light by light scattering graphs) appearing in deep inelastic 
lepton-hadron 
scattering (spacelike process). These graphs contribute in order $\alpha_s^3$ 
to $\hat {\cal F}_{3,q}^{\rm NS}$ but not to $\hat g_{1,q}^{\rm NS}$. 
In first order
of $\alpha_s$ the coefficient functions $\Delta {\cal C}_{4,q}^{\rm NS}$ and
$\Delta {\cal C}_{5,q}^{\rm NS}$ were calculated in \cite{ravin2} and they
satisfy the relations in Eq. (\ref{eqn:2.21}). The order $\alpha_s^2$ 
contributions are equal to ${\cal C}_{2,q}^{\rm NS}$ and 
${\cal C}_{1,q}^{\rm NS}$ respectively and can be found in \cite{zn1}.


\mysection{Order $\alpha_s^2$ contributions to the coefficient functions
$\Delta {\cal C}_{1,a}$}
In this section we present the results for the coefficient functions
$\Delta {\cal C}_{1,a}$ which emerge from the computation of the partonic
fragmentation functions $\hat g_{1,a}$ with $a=q,g$. The latter are collinearly
divergent and the singularities are regularized via n-dimensional 
regularization so that they manifest themselves as pole terms of the type
$(1/\varepsilon)^k$, with $\varepsilon=n-4$, in 
$\hat g_{1,a}(z,Q^2,\varepsilon)$. For the definition
of the $\gamma_5$-matrix we use the HVBM prescription \cite{hv}, \cite{bm}
( see also \cite{larin}, \cite{akde}).
This prescription destroys the anticommutativity of the $\gamma_5$-matrix in
Eq. (\ref{eqn:2.18}) so that $\hat g_{1,q}^{\rm NS} \not =
\hat {\cal F}_{3,q}^{\rm NS}$. This will lead to a violation of the relations
between the coefficients in Eq. (\ref{eqn:2.21}) and the nonrenormalization
properties of the non-singlet axial-vector current. Furthermore as is shown in
\cite{larin} the Adler-Bardeen theorem \cite{adba} will be also violated. 
These properties can be restored by introducing an additonal renormalization 
constant $Z_{qq}^{r}$ ($r= {\rm NS,S}$) which has the following unrenormalized
form
\begin{eqnarray}
\label{eqn:3.1}
Z_{qq}^{r}&=& \delta(1-z) + \frac{\hat \alpha_s}{4\pi} S_\varepsilon 
(\frac{M^2}{\mu^2})^{\varepsilon/2} \Big [ z_{qq}^{(1)}(z) +\varepsilon
b_{qq}^{(1)}(z) \Big ] 
\nonumber\\[2ex]
&& + (\frac{\hat \alpha_s}{4\pi})^2 S_\varepsilon^2 
(\frac{M^2}{\mu^2})^{\varepsilon}\Big [ -\frac{1}{\varepsilon} \beta_0 
z_{qq}^{(1)}(z) + z_{qq}^{r,(2)}(z) - 2 \beta_0 b_{qq}^{(1)}(z) \Big ]\,,
\end{eqnarray}
up to order $\alpha_s^2$. Further we have defined the spherical factor as
\begin{eqnarray}
\label{eqn:3.2}
S_\varepsilon = \exp\left[ \frac{1}{2}\varepsilon(\gamma_E - \ln 4\pi)\right]\,,
\end{eqnarray}
which is an artefact of n-dimensional regularization.
The same holds for the
scale $\mu$ which originates from the dimensionality of the gauge coupling
constant in n dimensions ($g \rightarrow g (\mu)^{-\varepsilon/2}$). The
mass parameter $M$ denotes the renormalization scale which will be put equal 
to the factorization scale.
Further  $\beta_0$ is the lowest-order coefficient in the series expansion of
the $\beta$-function which, up to order $g^5$, is given by
\begin{eqnarray}
\label{eqn:3.3}
\beta(g)=-\beta_0 \frac{g^3}{16\pi^2}-\beta_1\frac{g^5}{(16\pi^2)^2} 
\quad , \quad
\beta_0= \frac{11}{3}C_A-\frac{4}{3}n_fT_f ~~ ,~~ 
g^2\equiv 4\pi\alpha_s \,,
\end{eqnarray}
where $C_A=N$ and $T_f=1/2$ are colour factors in $SU(N)$. Notice that
in second order $z_{qq}^{{\rm S},(2)} = z_{qq}^{{\rm NS},(2)}
+n_f z_{qq}^{{\rm PS},(2)}$ 
where $n_f z_{qq}^{{\rm PS},(2)}$ is coming from the light by 
light scattering graphs in fig. 1 of \cite{larin}. In the non-singlet case
$Z_{qq}^{\rm NS}$ can be derived from
\begin{eqnarray}
\label{eqn:3.4}
Z_{qq}^{\rm NS}(z) = \frac{\hat g_{1,q}^{\rm NS}(z,Q^2,\mu^2)}
{\hat {\cal F}_{3,q}^{\rm NS}(z,Q^2,\mu^2)} \mid_{Q^2=M^2} \,.
\end{eqnarray}
The first order coefficients are given by
\begin{eqnarray}
\label{eqn:3.5}
z_{qq}^{(1)} = C_F \Big [ - 8 (1- z) \Big ]\,,
\end{eqnarray}
with $C_F=(N^2-1)/2N$ and
\begin{eqnarray}
\label{eqn:3.6}
b_{qq}^{(1)} = C_F \Big [- 8 (1- z) \ln z -  4 (1- z) \ln(1 - z) 
+ 10 - 8z \Big ]\,.
\end{eqnarray}
The second order coefficient $z_{qq}^{{\rm NS},(2)}$ is given in 
appendix A.
We will now present the general form of the partonic fragmentation functions
which follow from mass factorization and renormalization. The expressions 
are written in such a way that the non-logarithmic parts of the 
coefficient functions denoted by $\bar{c}_{1,k}$ ($k=q,g$) and the
the DGLAP splitting functions are presented in the $\overline{\rm MS}$-scheme.
This scheme holds for coupling constant renormalization as well as mass
factorization.

In the Born
approximation the partonic reaction in Eq. (\ref{eqn:2.15}) is given by 
(see fig. 2 in \cite{rn2})
\begin{eqnarray}
\label{eqn:3.7}
V \rightarrow q + \bar q \,,
\end{eqnarray}
where either the quark or the anti-quark is detected in the final state.
This leads to the contribution
\begin{eqnarray}
\label{eqn:3.8}
\hat g_{1,q}^{(0)} = 1 \equiv \delta(1-z) \,.
\end{eqnarray}
In order $\alpha_s$ we have the following contributions. First we have to
compute the one-loop virtual corrections to the Born reaction (\ref{eqn:3.7})
(see fig. 3 in \cite{rn2}) and add them to the gluon bremsstrahlung process
(see fig. 4 in \cite{rn2})
\begin{eqnarray}
\label{eqn:3.9}
V \rightarrow q + \bar q + g\,.
\end{eqnarray}
In the case the quark or the anti-quark is detected in the final state the
result becomes
\begin{eqnarray}
\label{eqn:3.10}
\hat g_{1,q}^{(1)} = \frac{\hat \alpha_s}{4\pi}\,S_\varepsilon\,
\left(\frac{Q^2}{\mu^2}\right)^{\varepsilon/2}\,\Big [ \Delta P_{qq}^{(0)}
 \frac{1}{\varepsilon} + \bar{c}_{1,q}^{(1)} + z_{qq}^{(1)}
+ \varepsilon\,\Big \{ \bar{a}_{1,q}^{(1)} + b_{qq}^{(1)} \Big \}\Big ]\,.
\end{eqnarray}
The timelike spin splitting function is given by
\begin{eqnarray}
\label{eqn:3.11}
\Delta P_{qq}^{(0)} = C_F \Big [8 \Big (\frac{1}{1-z} \Big )_{+} - 4 - 4z
+ 6 \delta(1-z) \Big ] \,,
\end{eqnarray}
and the non-pole terms are given by
\begin{eqnarray}
\label{eqn:3.12}
\bar{c}_{1,q}^{(1)}&=&C_F \Big [4 \Big (\frac{\ln(1-z)}{1-z}\Big )_+
- 3 \Big (\frac{1}{1-z} \Big )_+
\nonumber\\
&& - 2(1+z)\ln(1-z) + 4 \frac{1+z^2}{1-z} \ln(z) + 1 - z 
\nonumber\\[2ex]
&& + \delta(1-z) \Big ( -9 + 8 \zeta(2) \Big ) \Big] \,,
\end{eqnarray}
\begin{eqnarray}
\label{eqn:3.13}
\bar{a}_{1,q}^{(1)}&=&C_F \Big [ \Big (\frac{\ln^2(1-z)}{1-z}\Big )_+ -
\frac{3}{2}\Big (\frac{\ln(1-z)}{1-z}\Big )_+ + (\frac{7}{2}-3\zeta(2))
\Big (\frac{1}{1-z}\Big )_+
\nonumber\\
&& + 2\frac{1+z^2}{1-z} \Big (\ln(z)\ln(1-z) + \ln^2(z) \Big )
- 3\frac{\ln(z)}{1-z}
\nonumber\\
&& + (1+z) \Big ( \frac{3}{2}\zeta(2) - \frac{1}{2}\ln^2(1-z) \Big )
+ (1-z)\Big (\frac{1}{2}\ln(1-z)
\nonumber\\
&& + \ln(z)\Big ) - \frac{3}{2}+\frac{1}{2}z + \delta(1-z) \Big ( 9
- \frac{33}{4} \zeta(2) \Big ) \Big ] \,.
\end{eqnarray}
Notice that we also have to include in order $\alpha_s$ terms
proportional to $\varepsilon$, which are represented by $b_{qq}^{(1)}$  
Eq. (\ref{eqn:3.1}) and $\bar{a}_{1,k}^{(1)}$ ($k=q,g$)  
since they contribute via mass factorization to the non-logarithmic parts of
the coefficient functions.
When the gluon is detected in reaction (\ref{eqn:3.9}) we get
\begin{eqnarray}
\label{eqn:3.14}
\hat g_{1,g}^{(1)} = \frac{\hat \alpha_s}{4\pi}\,S_\varepsilon\,
\left(\frac{Q^2}{\mu^2}\right)^{\varepsilon/2}\,\Big [
2\Delta P_{gq}^{(0)}\frac{1}{\varepsilon} +
\bar{c}_{1,g}^{(1)} + \varepsilon\,\bar{a}_{1,g}^{(1)}\Big ]\,.
\end{eqnarray}
Here the timelike spin splitting function is given by
\begin{eqnarray}
\label{eqn:3.15}
\Delta P_{gq}^{(0)} = C_F \Big [ 8 - 4z \Big ] \,,
\end{eqnarray}
and the remaining coefficients are
\begin{eqnarray}
\label{eqn:3.16}
\bar{c}_{1,g}^{(1)}=C_F \Big [ (8 - 4 z) \Big ( \ln(1-z) + 2\ln(z) \Big ) - 16
+ 12z \Big ]  \,,
\end{eqnarray}
\begin{eqnarray}
\label{eqn:3.17}
\bar{a}_{1,g}^{(1)}&=&C_F \Big [ (2-z)\Big ( \ln^2(1-z) + 4\ln(z)\ln(1-z) +
4 \ln^2(z) \Big )
\nonumber\\
&& + (-8+6z)\Big ( \ln(1-z) + 2 \ln(z) - 2 \Big ) -(6-3z)\zeta(2) \Big ] \,.
\end{eqnarray}
In order $\alpha_s^2$ we have the following subprocesses. The non-singlet
partonic fragmentation function receives contributions from the two-loop
corrections to reaction (\ref{eqn:3.7}) (fig. 5 of \cite{rn2}), the
one-loop correction to reaction (\ref{eqn:3.9}) (fig. 6 of \cite{rn2}) and
the double gluon bremsstrahlung process (fig. 7 of \cite{rn2})
\begin{eqnarray}
\label{eqn:3.18}
V \rightarrow q + \bar q + g + g  \,,
\end{eqnarray}
where either the quark or the anti-quark is detected in the final state. Further
we have to add the contribution from the partonic subprocess
\begin{eqnarray}
\label{eqn:3.19}
V \rightarrow q + \bar q + q + \bar q \,,
\end{eqnarray}
with the condition that the two quarks or the two anti-quarks which belong to
the inclusive state are identical (see figs. 8,10 in \cite{rn2}). The result 
has the following form
\begin{eqnarray}
\label{eqn:3.20}
\hat{g}_{1,q}^{{\rm NS},(2)}&=& (\frac{\hat \alpha_s}{4\pi})^2\,
S_\varepsilon^2\,
\left(\frac{Q^2}{\mu^2}\right)^\varepsilon\,\Big [\frac{1}{\varepsilon^2}
\Big \{ \frac{1}{2} \Delta P_{qq}^{(0)}\otimes \Delta P_{qq}^{(0)} 
- \beta_0 \Delta P_{qq}^{(0)}\Big \}
\nonumber\\[2ex]
&& + \frac{1}{\varepsilon}\Big \{\frac{1}{2}\Big (\Delta P_{qq}^{{\rm NS},(1)} 
- \Delta P_{q\bar q}^{{\rm NS},(1)} \Big )
- 2\beta_0 \Big ( \bar{c}_{1,q}^{(1)} + z_{qq}^{(1)} \Big )
\nonumber\\[2ex]
&& +\Delta P_{qq}^{(0)}\otimes \Big ( \bar{c}_{1,q}^{(1)}+z_{qq}^{(1)}\Big ) 
\Big \}
 + \bar{c}_{1,q}^{{\rm NS},(2),{\rm nid}}-\bar{c}_{1,q}^{{\rm NS},(2),{\rm id}} 
+z_{qq}^{{\rm NS},(2)} 
\nonumber\\[2ex]
&&  + z_{qq}^{(1)} \otimes \bar{c}_{1,q}^{(1)} 
 - 2\beta_0 \Big (\bar{a}_{1,q}^{(1)}+ b_{qq}^{(1)} \Big ) 
+ \Delta P_{qq}^{(0)}\otimes \Big ( \bar{a}_{1,q}^{(1)}+
b_{qq}^{(1)}\Big )\Big ] \,.
\nonumber\\
\end{eqnarray}
The convolution symbol denoted by $\otimes$ is defined by
\begin{equation}
\label{eqn:3.21}
(f\otimes g)(z) = \int_0^1\,dz_1\,\int_0^1\,dz_2\,\delta(z-z_1z_2)f(z_1)g(z_2)
\,.
\end{equation}
The terms $\bar{c}_{1,q}^{(1)}$ and $\bar{a}_{1,q}^{(1)}$ already appeared in 
$\hat{g}_{1,q}^{(1)}$ (Eq. (\ref{eqn:3.10})). The order 
$\alpha_s^2$ non-singlet timelike spin splitting functions
$\Delta P_{qq}^{{\rm NS},(1)}$ and $\Delta P_{q\bar q}^{{\rm NS},(1)}$ are 
calculated in \cite{cfp} (see also \cite{stvo}) and can be found in Eqs.
(4.11) and (4.12) of \cite{rn2}. Notice that they are equal to the spin
averaged ones contrary to the singlet case which we will present below.
The splitting function $\Delta P_{q\bar q}^{{\rm NS},(1)}$ orginates from 
process 
(\ref{eqn:3.19}) with two identical (anti-)quarks in the inclusive state. The 
same holds for the non-pole term $\bar{c}_{1,q}^{{\rm NS},(2),{\rm id}}$ 
\footnote{ For the explicit expressions of 
$\bar{c}_{1,q}^{{\rm NS},(2),{\rm nid}}$ and 
$\bar{c}_{1,q}^{{\rm NS},(2),{\rm id}}$ see Eqs. (4.A.32) - (4.A.37)
in \cite{rijk}.}.
Notice that $\hat{g}_{1,q}^{{\rm NS},(2)}$ only appears if the amplitude
$G_{\mu\nu}$ (see Eq. \ref{eqn:2.18})) is projected on the non-singlet part.
If we project on the singlet part we get
\begin{eqnarray}
\label{eqn:3.22}
\hat{g}_{1,q}^{{\rm S},(2)}= \hat{g}_{1,q}^{{\rm NS},(2)}
+\hat{g}_{1,q}^{{\rm PS},(2)} \,,
\end{eqnarray}
with
\begin{eqnarray}
\label{eqn:3.23}
 \hat{g}_{1,q}^{{\rm PS},(2)} &=& n_f\,(\frac{\hat \alpha_s}{4\pi})^2\,
 S_\varepsilon^2\,
\left(\frac{Q^2}{\mu^2}\right)^\varepsilon\,\Big [\frac{1}{\varepsilon^2}
\Big \{\frac{1}{2} \Delta P_{gq}^{(0)}\otimes \Delta P_{qg}^{(0)}\Big \}
+ \frac{1}{\varepsilon}\Big \{\frac{1}{2} \Delta P_{qq}^{{\rm PS},(1)}
\nonumber\\[2ex]
&& +\frac{1}{2}\Delta P_{qg}^{(0)}\otimes\bar{c}_{1,g}^{(1)}\Big \} +
\bar{c}_{1,q}^{{\rm PS},(2)} + \frac{1}{2}\Delta P_{qg}^{(0)}\otimes
\bar{a}_{1,g}^{(1)} + z_{qq}^{{\rm PS},(2)} \Big ]\,,
\end{eqnarray}
where $\bar{c}_{1,g}^{(1)}$ and $\bar{a}_{1,g}^{(1)}$ are given in Eq. 
(\ref{eqn:3.14}).
The purely singlet (PS) part originates from process (\ref{eqn:3.19}) which
proceeds by the exchange of two gluons in the t-channel (see figs. 8,9 in 
\cite{rn2}). Further $z_{qq}^{{\rm S},(2)}= z_{qq}^{{\rm NS},(2)}
+n_f z_{qq}^{{\rm PS},(2)}$ where $n_f z_{qq}^{{\rm PS},(2)}$ originates from
the light by light scattering graphs in fig. 1 in \cite{larin}. In the
second order expression above the lowest order splitting function 
$\Delta P_{qg}^{(0)}$ appears for the first time and it reads
\begin{eqnarray}
\label{eqn:3.24}
\Delta  P_{qg}^{(0)} = T_f (-8 + 16z) \,.
\end{eqnarray}
From our calculations and the expression given in Eq. (\ref{eqn:3.23}) one can 
also infer the splitting function 
\begin{eqnarray}
\label{eqn:3.25}
\Delta P_{qq}^{{\rm PS},(1)} = C_FT_f \Big[ (16 + 16 z) \ln^2 z
- (48 + 112 z)\ln z -176 + 176 z \Big] \,.
\end{eqnarray}
The expression above agrees with the recent result 
obtained in \cite{stvo}. The complete
singlet spin timelike splitting function is then given by
\begin{eqnarray}
\label{eqn:26}
\Delta P_{qq}^{{\rm S},(1)}=\Delta P_{qq}^{{\rm NS},(1)}
-\Delta P_{q\bar q}^{{\rm NS},(1)}+\Delta P_{qq}^{{\rm PS},(1)} \,.
\end{eqnarray}
If the gluon is detected in the final state the one-loop corrections to
reaction (\ref{eqn:3.9}) and the contribution due to the gluon bremsstrahlung
process in Eq. (\ref{eqn:3.18}) lead to the answer.
\begin{eqnarray}
\label{eqn:3.27}
\hat{g}_{1,g}^{(2)}&=& (\frac{\hat \alpha_s}{4\pi})^2\,
S_\varepsilon^2\,
\left(\frac{Q^2}{\mu^2}\right)^\varepsilon\,\Big [\frac{1}{\varepsilon^2}
 \Big \{ \Delta P_{gq}^{(0)}\otimes(\Delta P_{gg}^{(0)} + \Delta P_{qq}^{(0)})
 - 2\beta_0 \Delta P_{gq}^{(0)}\Big \}
\nonumber\\[2ex]
  && + \frac{1}{\varepsilon}\Big \{ \Big ( \Delta P_{gq}^{(1)} 
- \Delta P_{gq}^{(0)} \otimes z_{qq}^{(1)} \Big ) - 2\beta_0\bar{c}_{1,g}^{(1)}
  +\Delta P_{gg}^{(0)}\otimes\bar{c}_{1,g}^{(1)} 
\nonumber\\[2ex]
&&+ 2 \Delta P_{gq}^{(0)}\otimes \Big ( \bar{c}_{1,q}^{(1)} + z_{qq}^{(1)} 
\Big ) \Big \}
 + \bar{c}_{1,g}^{(2)} - 2\beta_0 \bar{a}_{1,g}^{(1)} 
 + \Delta P_{gg}^{(0)}\otimes \bar{a}_{1,g}^{(1)}
\nonumber\\[2ex]
&& + 2 \Delta P_{gq}^{(0)}\otimes 
\Big (\bar{a}_{1,q}^{(1)} + b_{qq}^{(1)} \Big )\Big ] \,.
\end{eqnarray}
In the second order expression above one encounters for the first time the 
lowest order splitting function
\begin{eqnarray}
\label{eqn:3.28}
\Delta P_{gg}^{(0)} = C_A \left(8 \Big (\frac{1}{1-z} \Big)_+ + 8 - 16z
+ \frac{22}{3}\delta(1-z)\right) - \frac{8}{3}n_fT_f\delta(1-z)\,.
\end{eqnarray}
From our calculation we also infer the second order spin timelike splitting
function
\begin{eqnarray}
\label{eqn:3.29}
\Delta  P_{gq}^{(1)} &=& C_F^2\Big [ (128 - 64z)\Big ( {\rm Li}_2(1-z)
+ \frac{1}{2}\ln z\ln(1-z)
\nonumber\\[2ex]
&& -\frac{1}{16}\ln^2z + \frac{1}{8}\ln^2(1-z)-\zeta(2)\Big ) +(64 - 36z)\ln z
\nonumber\\[2ex]
&& - 32(1-z)\ln(1-z) + 76 - 56z\Big ]
\nonumber\\[2ex]
&& +C_AC_F\Big [ (32 + 16z)\Big ({\rm Li}_2(-z)+\ln z\ln(1+z)\Big ) 
+ (-128 + 64z)
\nonumber\\[2ex]
&&\times \Big ({\rm Li}_2(1-z) + \frac{1}{4}\ln z\ln(1-z) 
+\frac{1}{8}\ln^2(1-z) \Big )
- (48 + 8z)\ln^2z
\nonumber\\[2ex]
&& + (128 - 48z)\zeta(2) + (-32 + 88z)\ln z + 32(1-z)\ln(1-z)
\nonumber\\[2ex]
&& + 40 -32z \Big ] \,,
\end{eqnarray}
which agrees with the recent result in \cite{stvo}.\\
To obtain the finite coefficient functions in the $\overline{\rm MS}$-scheme
one has to perform mass factorization. The relations between the partonic
fragmentation functions and the coefficient functions are then given by
\begin{eqnarray}
\label{eqn:3.30}
\hat g_{1,q}^{\rm NS} = Z_{qq}^{\rm NS}\otimes 
\overline{\Gamma}_{qq}^{\rm NS}\otimes
\Delta \overline{{\cal C}}_{1,q}^{\rm NS} \,,
\end{eqnarray}
\begin{eqnarray}
\label{eqn:3.31}
\hat g_{1,q}^{\rm S} = Z_{qq}^{\rm S}\otimes 
\overline{\Gamma}_{qq}^{\rm S}\otimes
\Delta \overline{{\cal C}}_{1,q}^{\rm S} + n_f 
\overline{\Gamma}_{qg} \otimes \Delta \overline{{\cal C}}_{1,g} \,,
\end{eqnarray}
\begin{eqnarray}
\label{eqn:3.32}
\hat g_{1,g} = 2 Z_{qq}^{\rm S}\otimes \overline{\Gamma}_{gq} \otimes 
\Delta \overline{{\cal C}}_{1,q}^{\rm S}
+ \overline{\Gamma}_{gg} \otimes \Delta \overline{{\cal C}}_{1,g} \,.
\end{eqnarray}
Here the renormalization factor $Z_{qq}^r$ is needed to restore the 
anticommutativity of the $\gamma_5$-matrix on the level of the coefficient
functions so that they are presented in the genuine $\overline{\rm MS}$-scheme.
If we expand the transition functions in the unrenormalised coupling constant
$\bar \alpha_s$ they read as follows
\begin{eqnarray}
\label{eqn:3.33}
\overline{\Gamma}_{qq}^{\rm NS}&=&1+\frac{\hat \alpha_s}{4\pi}\,S_\varepsilon\,
\left(\frac{M^2}{\mu^2}\right)
^{\varepsilon/2}\Big [\frac{1}{\varepsilon} \Delta P_{qq}^{(0)}\Big ]
+ (\frac{\hat \alpha_s}{4\pi})^2\,S_\varepsilon^2\,\left(
\frac{M^2}{\mu^2}\right)^\varepsilon
\nonumber\\[2ex]
&&\times \Big [ \frac{1}{\varepsilon^2}
  \Big \{ \frac{1}{2} \Delta P_{qq}^{(0)} \otimes \Delta P_{qq}^{(0)}
-\beta_0 \Delta P_{qq}^{(0)}\Big \} 
\nonumber\\[2ex]
&&+ \frac{1}{\varepsilon} \Big \{\frac{1}{2}\Big ( \Delta P_{qq}^{{\rm NS},(1)}
- \Delta P_{q\bar{q}}^{{\rm NS},(1)}\Big ) \Big \} \Big ] \,.
\end{eqnarray}
The singlet transition function is given by
\begin{eqnarray}
\label{eqn:3.34}
\overline{\Gamma}_{qq}^{\rm S} = \overline{\Gamma}_{qq}^{\rm NS}
+ 2n_f\,\overline{\Gamma}_{qq}^{\rm PS} \,,
\end{eqnarray}
where the purely singlet (PS) part can be written as
\begin{eqnarray}
\label{eqn:3.35}
\overline{\Gamma}_{qq}^{\rm PS} = (\frac{\hat \alpha_s}{4\pi})^2\,
S_\varepsilon^2\,
\left(\frac{M^2}{\mu^2}\right)^\varepsilon\,\Big [\frac{1}{\varepsilon^2}
\Big \{\frac{1}{4} \Delta P_{gq}^{(0)}\otimes \Delta P_{qg}^{(0)}\Big \} + 
\frac{1}{\varepsilon}
\Big \{\frac{1}{4} \Delta P_{qq}^{{\rm PS},(1)}\Big \}\Big ] \,.
\end{eqnarray}
Finally we have
\begin{eqnarray}
\label{eqn:3.36}
\overline{\Gamma}_{gq} &=& \frac{\hat \alpha_s}{4\pi}\,S_\varepsilon\,\left(
\frac{M^2}{\mu^2}\right)^{\varepsilon/2}\,
\Big [\frac{1}{\varepsilon} \Delta P_{gq}^{(0)}\Big ]
+(\frac{\hat \alpha_s}{4\pi})^2\,S_\varepsilon^2\,\left(\frac{M^2}{\mu^2}\right)
^\varepsilon 
\nonumber\\[2ex]
&& \times \Big [\frac{1}{\varepsilon^2}\Big \{\frac{1}{2} 
\Delta P_{gq}^{(0)} \otimes
(\Delta P_{gg}^{(0)}+ \Delta P_{qq}^{(0)})-\beta_0 \Delta P_{gq}^{(0)}\Big \} 
\nonumber\\[2ex]
&& + \frac{1}{\varepsilon}
\Big \{\frac{1}{2}\Big ( \Delta P_{gq}^{(1)} - \Delta P_{gq}^{(0)} \otimes
z_{qq}^{(1)} \Big ) \Big \} \Big ] \,.
\end{eqnarray}
We only need the following transition functions up to order $\hat \alpha_s$
\begin{eqnarray}
\label{eqn:3.37}
\overline{\Gamma}_{qg} = \frac{\hat \alpha_s}{4\pi}\,S_\varepsilon\,\left(
\frac{M^2}{\mu^2}\right)^{\varepsilon/2}\,
\Big [\frac{1}{2\varepsilon}\Delta P_{qg}^{(0)}\Big ] \,,
\end{eqnarray}
\begin{eqnarray}
\label{eqn:3.38}
\overline{\Gamma}_{gg} = 1 + \frac{\hat \alpha_s}{4\pi}\,S_\varepsilon\,
\left(\frac{M^2}{\mu^2}\right)^{\varepsilon/2}\,
\Big [\frac{1}{\varepsilon}\Delta P_{gg}^{(0)}\Big ] \,.
\end{eqnarray}
Further we have to perform coupling constant renormalization which is also
carried out in the $\overline{\rm MS}$-scheme. Up to order $\alpha_s^2$ it
is sufficient to replace the bare coupling constant by
\begin{eqnarray}
\label{eqn:3.39}
\frac{\hat \alpha_s}{4\pi} = \frac{\alpha_s(M^2)}{4\pi} \Big ( 1 +
\frac{\alpha_s(M^2)}{4\pi} \frac{2\beta_0}{\varepsilon} S_\varepsilon
\left(\frac{M^2}{\mu^2}\right)^{\varepsilon/2} \Big )\,.
\end{eqnarray}
The coefficient functions ($\overline{\rm MS}$-scheme) have the following 
representations
\begin{eqnarray}
\label{eqn:3.40}
\Delta \overline{{\cal C}}_{1,q}^{\rm NS} &=& 1 + \frac{\alpha_s}{4\pi}
\Big [\frac{1}{2} \Delta P_{qq}^{(0)} L_M
+ \bar{c}_{1,q}^{(1)}\Big ] + (\frac{\alpha_s}{4\pi})^2\,\Big [
\Big \{\frac{1}{8} \Delta P_{qq}^{(0)}\otimes  \Delta P_{qq}^{(0)}
\nonumber\\[2ex]
&& - \frac{1}{4}\beta_0 \Delta P_{qq}^{(0)}
\Big \} L_M^2 + \Big \{\frac{1}{2}(\Delta P_{qq}^{{\rm NS},(1)}
-\Delta P_{q\bar{q}}^{{\rm NS},(1)}) -\beta_0\bar{c}_{1,q}^{(1)}
\nonumber\\[2ex]
&& + \frac{1}{2} \Delta P_{qq}^{(0)}\otimes\bar{c}_{1,q}^{(1)} \Big \} L_M
+ \bar{c}_{1,q}^{{\rm NS},(2),{\rm nid}}
- \bar{c}_{1,q}^{{\rm NS},(2),{\rm id}} \Big ] \,,
\end{eqnarray}
\begin{eqnarray}
\label{eqn:3.41}
\Delta \overline{{\cal C}}_{1,q}^{\rm PS} &=& n_f\,(\frac{\alpha_s}{4\pi})^2\,
\Big [ \Big \{\frac{1}{8} \Delta P_{gq}^{(0)}\otimes P_{qg}^{(0)}\Big \} L_M^2 
\nonumber\\[2ex]
&& + \Big \{\frac{1}{2} \Delta P_{qq}^{{\rm PS},(1)} + \frac{1}{4} 
\Delta P_{qg}^{(0)} \otimes\bar{c}_{1,g}^{(1)} \Big \} L_M
+ \bar{c}_{1,q}^{{\rm PS},(2)}\Big ] \,,
\end{eqnarray}
\begin{eqnarray}
\label{eqn:3.42}
\Delta \overline{{\cal C}}_{1,g} &=& \frac{\alpha_s}{4\pi}\,\Big [ 
\Delta P_{gq}^{(0)} L_M
 + \bar{c}_{1,g}^{(1)}\Big ] + (\frac{\alpha_s}{4\pi})^2 \,
\Big [ \Big \{\frac{1}{4} \Delta P_{gq}^{(0)}\otimes( \Delta P_{gg}^{(0)}
+ \Delta P_{qq}^{(0)})
\nonumber\\[2ex]
&& - \frac{1}{2}\beta_0 \Delta P_{gq}^{(0)}\Big \} L_M^2 + \Big \{ 
\Delta P_{gq}^{(1)}
-\beta_0\bar{c}_{1,g}^{(1)} + \frac{1}{2} \Delta P_{gg}^{(0)}\otimes
\bar{c}_{1,g}^{(1)}
\nonumber\\[2ex]
&& + \Delta P_{gq}^{(0)}\otimes\bar{c}_{1,q}^{(1)}\Big \} L_M 
+ \bar{c}_{1,g}^{(2)}\Big ] \,,
\end{eqnarray}
with the definitions
\begin{eqnarray}
\label{eqn:3.43}
L_M = \ln\frac{Q^2}{M^2} \quad , \quad \alpha_s \equiv \alpha_s(M^2) \,.
\end{eqnarray}
In the above expressions $M$ denotes the factorization scale which has been put
equal to the renormalization scale defined by $R$. In the case these scales
are chosen to be different the corresponding coefficient functions can be
derived from expressions (\ref{eqn:3.41})-(\ref{eqn:3.43}) by the replacement
\begin{eqnarray}
\label{eqn:3.44}
\alpha_s(M^2) = \alpha_s(R^2)\left[ 1+\frac{\alpha_s(R^2)}{4\pi}\,\beta_0\,
\ln\frac{R^2}{M^2}\right].
\end{eqnarray}
As has been indicated in Eq. (\ref{eqn:2.21}) 
$\Delta \overline{{\cal C}}_{1,q}^{\rm NS}$ 
(\ref{eqn:3.40}) is equal to $\overline{{\cal C}}_{3,q}^{\rm NS}$
and the latter is calculated in \cite{rn3} (see
also \cite{rijk} ). Since $z_{qq}^{{\rm PS},(2)}$ is not known yet, except for 
the first moment (see below), we can only present $\bar{c}_{1,q}^{{\rm PS},(2)}
+ z_{qq}^{{\rm PS},(2)}$ in ${\cal C}_{1,q}^{\rm PS}$ (see Eqs.
(\ref{eqn:3.41}), (\ref{eqn:A.1}) ) but not $\bar{c}_{1,q}^{{\rm PS},(2)}$
and $z_{qq}^{{\rm PS},(2)}$ separately. Unfortunately we cannot infer
the latter from a ratio of two partonic fragmentation functions
like we could do for $z_{qq}^{{\rm NS},(2)}$ in Eq. (\ref{eqn:3.4}). In the 
purely singlet case one has to compute the three-loop operator matrix
elements for general moments (spin) given by the light by light scattering 
graphs in fig. 1 of \cite{larin}. The renormalization of these graphs is ruled
by the Adler-Bardeen theorem \cite{adba} which is hard to impose on the level
of the second order partonic fragmentation functions. For that one has to 
calculate the latter quantities in third order which is a tantalizing 
enterprise.
Fortunately this does not hamper the computation of 
$\Delta \overline{{\cal C}}_{1,g}$ 
(\ref{eqn:3.42})
up to second order, in which $z_{qq}^{{\rm PS},(2)}$ does not show up, and the
result can be found in Eq. (\ref{eqn:A.2}).

Finally we want to mention another problem which concerns the 
renormalization constant $Z_{qq}^r$ ($r={\rm NS,S}$) in Eq. (\ref{eqn:3.1})
which is needed to restore the Ward-identities broken by the HVBM-prescription 
for the $\gamma_5$-matrix. After coupling constant renormalization it reads
\begin{eqnarray}
\label{eqn:3.45}
Z_{qq}^{r}&=& \delta(1-z) + \frac{\alpha_s}{4\pi} S_\varepsilon
(\frac{M^2}{\mu^2})^{\varepsilon/2} \Big [ z_{qq}^{(1)}(z) +\varepsilon
b_{qq}^{(1)}(z) \Big ]
\nonumber\\[2ex]
&& + (\frac{\alpha_s}{4\pi})^2 S_\varepsilon^2
(\frac{M^2}{\mu^2})^{\varepsilon}\Big [ \frac{1}{\varepsilon} \beta_0
z_{qq}^{(1)}(z) + z_{qq}^{r,(2)}(z) \Big ] \,,
\end{eqnarray}
where $z_{qq}^{(1)}$ and $b_{qq}^{(1)}$ are given in Eqs. (\ref{eqn:3.5}) and
(\ref{eqn:3.6}) respectively. In the non-singlet case $z_{qq}^{{\rm NS},(2)}$
can be inferred from Eq. (\ref{eqn:3.4}) and it is presented in 
Eq. (\ref{eqn:A.3}). The first moment of Eq. (\ref{eqn:3.45}) is equal to
\begin{eqnarray}
\label{eqn:3.46}
Z_{qq}^{\rm NS}&=& 1 + \frac{\alpha_s}{4\pi} S_\varepsilon C_F \Big [ - 4  
+ 13 \varepsilon \Big ] + (\frac{\alpha_s}{4\pi})^2 S_\varepsilon^2 \Big [
C_F^2 \{ 12 + 16 \zeta(2)\} 
\nonumber\\[2ex]
&& + C_A C_F \{ -\frac{44}{3} \frac{1}{\varepsilon}
- \frac{107}{9} \} + n_f C_F T_f \{ \frac{16}{3} 
\frac{1}{\varepsilon} + \frac{4}{9} \} \Big ] \,,
\end{eqnarray}
where we have put $M=\mu$. Comparing the result above with the one obtained
in \cite{lave} given by
\begin{eqnarray}
\label{eqn:3.47}
Z_{qq}^{\rm NS}&=& 1 + \frac{\alpha_s}{4\pi} S_\varepsilon C_F \Big [ - 4 
+ 5 \varepsilon \Big ] + (\frac{\alpha_s}{4\pi})^2 S_\varepsilon^2 \Big [
C_F^2 \{ 22 \} 
\nonumber\\[2ex]
&& + C_A C_F \{ -\frac{44}{3} \frac{1}{\varepsilon}
- \frac{107}{9} \} + n_f C_F T_f \{ \frac{16}{3} 
\frac{1}{\varepsilon} + \frac{4}{9} \} \Big ] \,,
\end{eqnarray}
we observe a discrepancy which shows up in the order $\alpha_s^2$ coefficient
of the colour factor $C_F^2$. Notice that the result in Eqs.
(8), (11) of \cite{lave} is obtained by computing the ratio between the 
vector and the axial-vector form factor where it was needed to restore the 
Ward-identity for the non-singlet axial vector current broken by the
HVBM-prescription. A discrepancy is also oberved when Eq. (\ref{eqn:3.4}) is
computed for deep inelastic scattering where $q^2=-Q^2$ (see \cite{zn3}).
Here one gets an answer which differs from the result in Eq. (\ref{eqn:3.46})
but agrees with Eq. (\ref{eqn:3.47}).
From the above we conclude that the renormalization constant $Z_{qq}^{\rm NS}$ 
is process dependent except for leading order.\\ 
The difference of this constant between deep inelastic structure functions 
(spacelike process) and the fragmentation functions (timelike process) is of 
the same origin as the one discovered for spacelike and timelike splitting 
functions in \cite{cfp}. The process dependence of the renormalization constant 
above is in contrast 
to what is known about the usual
renormalization constants computed in cases where the regularization 
procedure does not violate the Ward-identities.
In \cite{larin} also the first moment of $Z_{qq}^{\rm S}$
has been computed. The difference between $Z_{qq}^{\rm NS}$ and 
$Z_{qq}^{\rm S}$ only shows up in the $n_fC_FT_f$-part of Eq. (\ref{eqn:3.47}).
In the singlet case one has to add to the latter expression the first 
moment of $n_f z_{qq}^{{\rm PS},(2)}$ which equals $3n_fC_FT_f$. 
Since the leading terms in $Z_{qq}^r$ ($r={\rm NS,S}$)
are process independent we can assume that this also holds for 
$z_{qq}^{{\rm PS},(2)}$ so that at least we know the first moment for
$\Delta \overline{{\cal C}}_{1,q}^{\rm PS}$ (Eq. (\ref{eqn:3.41})).\\

Summarizing the above we have computed the order $\alpha_s^2$ contributions
to the coefficient functions corresponding to the spin fragmentation functions
$g_i^{\rm H}(x,Q^2)$ ($i=1,4,5$) measured in electron-positron annihilation
in which the hadron H is polarized. In addition we obtained the second
order splitting functions $\Delta P_{gq}^{{\rm PS},(1)}$, $\Delta P_{gq}^{(1)}$
which agree with the recent results published in \cite{stvo}. Unfortunately
we could not get the remaining ones given by $P_{qg}^{(1)}$ and $P_{qg}^{(1)}$
which could be calculated by the cut vertex method in \cite{stvo}. 
This is because the splitting functions $P_{qg}^{(i)}$, $P_{gg}^{(i)}$ always 
appear one order higher in the coefficient functions than $P_{qq}^{(i)}$,
$P_{gq}^{(i)}$. This property can be traced back to the fact that the
electroweak vector bosons never directly couple to the gluon. Finally we 
want to mention that for a complete next-to-next-to-leading order analysis
one also needs the three-loop splitting functions which have to be combined
with the coefficient functions in Eqs. (\ref{eqn:3.40})-(\ref{eqn:3.42}).
The computation of these splitting functions for spacelike and timelike
processes is one of the outstanding enterprises which have to be still carried
out.\\[5mm]
\noindent
Acknowledgements\\

We would like to thank J. Smith for the critical reading of
the manuscript and for giving us some useful comments.


\appendix
\mysection*{Appendix A}
\setcounter{section}{1}
In this section we only present the coefficient functions 
$\Delta \overline{{\cal C}}_{1,q}^{\rm PS}$
and $\Delta \overline{{\cal C}}_{1,g}$ since the non-singlet ones, given by  
$\Delta \overline{{\cal C}}_{i,q}^{\rm NS}$
($i=1,4,5$), are already known in the literature via the relations in 
Eq. (\ref{eqn:2.21}). The purely singlet coefficient equals
\begin{eqnarray}
\label{eqn:A.1}
\Delta \overline{{\cal C}}_{1,q}^{\rm PS} &=& n_f(\frac{\alpha_s}{4\pi})^2 
C_FT_f \Big [ \Big \{ 8(1+z)\ln z + 20(1-z) \Big \} L_M^2 
\nonumber\\[2ex]
&& + \Big \{ 16(1 +z)\Big ({\rm Li}_2(1-z) + \ln z\ln(1-z) \Big ) 
+ 24(1+z)\ln^2z  
\nonumber\\[2ex]
&& +(8-88z)\ln z + 40(1-z)\ln(1-z) - 136(1-z) \Big \} L_M 
\nonumber\\[2ex]
&& +16(1+z) \Big ( 3{\rm S}_{1,2}(1-z)-{\rm Li}_3(1-z) + 3\ln z {\rm Li}_2(1-z) 
\nonumber\\[2ex]
&& + \ln(1-z) {\rm Li}_2(1-z) + \frac{1}{2}\ln z\ln^2(1-z) + 
\frac{3}{2}\ln^2z\ln(1-z) 
\nonumber\\[2ex]
&& + \frac{11}{12}\ln^3z 
 - \zeta(2)\ln z \Big ) + (8 - 88z)\Big ( {\rm  Li}_2(1-z)
 + \ln z\ln(1-z) \Big ) 
\nonumber\\[2ex]
&& - \Big (\frac{32}{3z} 
 + 32 + 32z + \frac{32}{3}z^2\Big )\Big ( {\rm Li}_2(-z) + \ln z\ln(1+z) \Big )
  + \Big ( -26 
\nonumber\\[2ex]
&&- 58z + \frac{16}{3}z^2\Big ) \ln^2z + 20(1-z)\ln^2(1-z)
  + \Big (-72 + 40z 
\nonumber\\[2ex]
&& - \frac{32}{3}z^2\Big )\zeta(2) + \left(-\frac{364}{3} + \frac{356}{3}z
  \right)\ln z - 136(1-z)\ln(1-z) 
\nonumber\\[2ex]
  && + \frac{472}{3}(1-z)\Big ] 
- n_f (\frac{\alpha_s}{4\pi})^2 z_{qq}^{{\rm PS},(2)}(z) \,,
\end{eqnarray}
with
\begin{eqnarray}
\label{eqn:A.2}
z_{qq}^{{\rm PS},(2)}(z)=  C_F T_f \Big [ - 4 (1 + 2 z) \ln^2 z 
+ 8 (- 1 + 3 z) \ln z + 16 ( 1 - z) \Big ] \,.
\end{eqnarray}
The gluonic coefficient function becomes
\begin{eqnarray}
\label{eqn:A.3}
\Delta \overline{{\cal C}}_{1,g}^{(2)} &=& (\frac{\alpha_s}{4\pi})^2 
\Big [C_F^2 \Big \{ \Big ( (-8 + 4z)\ln z + (16 - 8z)\ln(1-z)
\nonumber\\[2ex]
&& + 6z\Big ) L_M^2 + \Big ( (48 - 24z) {\rm Li}_2(1-z) + (-24 + 12z)\ln^2z
  + (48 
\nonumber\\[2ex]
&& - 24z) \ln z\ln(1-z) + (32 - 16z)\ln^2(1-z) + (-64 + 32z)\zeta(2)
\nonumber\\[2ex]
&&+ (128 + 4z)\ln z + (-80 + 68z)\ln(1-z) + 132 - 148z\Big ) L_M
\nonumber\\[2ex]
&& + \Big (-\frac{96}{z} + 64 - 96z\Big ) {\rm Li}_3(-z)+\Big (\frac{64}{z}  
+ 128 + 64z \Big ) {\rm S}_{1,2}(-z)
\nonumber\\[2ex]
&&+\Big (\frac{32}{z}-240 + 120z\Big ) {\rm S}_{1,2}(1-z) + \Big (\frac{64}{z} 
+ 128 + 64z\Big )
\nonumber\\[2ex]
&&\times \ln(1+z) {\rm Li}_2(-z) + (-48 + 24z)\ln z {\rm Li}_2(1-z) 
+ (32 - 16z)
\nonumber\\[2ex]
&&\times \ln(1-z) {\rm Li}_2(1-z) + \Big ( \frac{32}{z} - 64 + 32z\Big )\ln z 
{\rm Li}_2(-z)+ (8 - 4z)
\nonumber\\[2ex]
&&\times \ln^2z\ln(1-z) - \Big (\frac{16}{z} + 32 + 16z\Big )\ln^2z\ln(1+z)
  + (40 - 20z)
\nonumber\\[2ex]
&&\times \ln z\ln^2(1-z) + \Big (\frac{32}{z} + 64 + 32z\Big )\ln z\ln^2(1+z)
  + \Big ( -\frac{44}{3} 
\nonumber\\[2ex]
&&+ \frac{22}{3}z\Big )\ln^3z + \Big (\frac{40}{3} - \frac{20}{3}z\Big )
\ln^3(1-z)+ (16 - 8z)\zeta(2)\ln z
\nonumber\\[2ex]
&& + \Big ( -\frac{32}{z} + 48 - 24z\Big )\zeta(2)\ln(1-z) + \Big ( 
\frac{32}{z} + 64 + 32z\Big )\zeta(2)
\nonumber\\[2ex]
&&\times \ln(1+z) + \Big ( -\frac{32}{z} + 80 - 104z\Big )\zeta(3) 
+ (-32 + 20z) {\rm Li}_2(1-z)
\nonumber\\[2ex]
&& + (-64 + 52 z) \ln z\ln(1-z) - \Big ( \frac{208}{3z} + 64 
+ \frac{64}{3}z^2\Big ) \Big ( {\rm Li}_2(-z)
\nonumber\\[2ex]
&& + \ln z\ln(1+z)\Big ) + \Big ( 128 - 43z + \frac{32}{3}z^2\Big )\ln^2z
  + (-72 + 54z)
\nonumber\\[2ex]
&&\times \ln^2(1-z) + \Big ( 32 - 60z - \frac{64}{3}z^2\Big )\zeta(2)
  + \Big (\frac{94}{3} + \frac{196}{3}z \Big )\ln z 
\nonumber\\[2ex]
&& + \Big (\frac{16}{z} + 88 - 108z\Big )\ln(1-z) + \frac{98}{3}
  - \frac{110}{3}z \Big \}
\nonumber\\[2ex]
&& +C_AC_F \Big \{ \Big ( -(32+8z)\ln z + (16 - 8z)\ln(1-z) - 48(1-z)\Big )
\nonumber\\[2ex]
&&\times L_M^2 + \Big ( -96 {\rm Li}_2(1-z) - 32(1+z)\ln z\ln(1-z) 
+ (32 + 16z)
\nonumber\\[2ex]
&& \times \Big ( {\rm Li}_2(-z) + \ln z\ln(1+z)\Big ) - (112 + 24z)\ln^2z 
+ (16 - 8z)
\nonumber\\[2ex]
&&\times \ln^2(1-z) + (96 - 32z)\zeta(2) + (-32 + 136z)\ln z + (-128 
\nonumber\\[2ex]
&& + 112z)\ln(1-z) + 232 - 224z\Big ) L_M
  + (32 + 16z)\Big ( {\rm Li}_3 \left(-\frac{1-z}{1+z}\right) 
\nonumber\\[2ex]
&&-{\rm Li_3} \left(\frac{1-z}{1+z}\right)\Big ) + (80 + 40z){\rm Li}_3(1-z)
  - \Big (\frac{16}{z} + 320\Big )
\nonumber\\[2ex]
&&\times {\rm S}_{1,2}(1-z)+\Big (\frac{48}{z} + 32 + 16z\Big ) {\rm Li}_3(-z) 
- \Big (\frac{32}{z}+ 64 + 32z\Big )
\nonumber\\[2ex]
&&\times {\rm S}_{1,2}(-z)-(288 + 16z)\ln z {\rm Li}_2(1-z) - (80 + 8z)\ln(1-z)
\nonumber\\[2ex]
&&\times {\rm Li}_2(1-z) + (32 + 16z)\ln(1-z) {\rm Li}_2(-z) 
+ \Big ( -\frac{16}{z} + 32 + 16z\Big )
\nonumber\\[2ex]
&&\times \ln z {\rm Li}_2(-z) - \Big (\frac{32}{z} + 64 + 32z\Big )\ln(1+z) 
{\rm Li}_2(-z) - \Big (\frac{16}{z} + 32
\nonumber\\[2ex]
&& + 16z\Big )\ln z\ln^2(1+z) - (96 + 32z)\ln^2z\ln(1-z) - \Big (\frac{248}{3}
\nonumber\\[2ex]
&& + \frac{44}{3}z\Big )\ln^3z + (32 + 16z)\ln z\ln(1-z)\ln(1+z) - (24 + 12z)
\nonumber\\[2ex]
&&\times \ln z\ln^2(1-z) + \Big (\frac{8}{z} + 48 + 24z\Big )\ln^2z\ln(1+z) + 
\Big (  \frac{8}{3} - \frac{4}{3}z\Big )
\nonumber\\[2ex]
&&\times \ln^3(1-z) + (256 - 48z)\zeta(2)\ln z + \Big (\frac{16}{z} + 16z\Big )
\zeta(2)\ln(1-z)
\nonumber\\[2ex]
&& - \Big (\frac{16}{z} + 32 + 16z\Big )\zeta(2)\ln(1+z) + \Big (
\frac{16}{z} + 24 + 20z\Big )\zeta(3)
\nonumber\\[2ex]
&& + (64 + 56z) {\rm Li}_2(1-z) + \Big ( 72 + 38z - \frac{16}{3}z^2\Big )
\ln^2z + 
\nonumber\\[2ex]
&& (-56 + 48z)\ln^2(1-z) + \Big (\frac{176}{3z} + 64 + 32z + \frac{32}{3}z^2
\Big )\Big ( {\rm Li}_2(-z)
\nonumber\\[2ex]
&& + \ln z\ln(1+z)\Big ) + (-48 + 136z)\ln z\ln(1-z) + \Big ( 48 - 16z 
\nonumber\\[2ex]
&& + \frac{32}{3}z^2\Big )\zeta(2) + \Big (\frac{844}{3} - \frac{128}{3}z\Big )
\ln z  + \Big (-\frac{8}{z} + 276 - 240z\Big )
\nonumber\\[2ex]
&& \times \ln(1-z) - \frac{124}{3} + \frac{76}{3}z\Big \}\Big ]\,.
\end{eqnarray}
The renormalization constant $Z_{qq}^{\rm NS}$ in Eq. (\ref{eqn:3.46}) 
needed to restore the Ward-identities, which are broken
by the HVBM prescription of the $\gamma_5$-matrix, is equal to 
\begin{eqnarray}
\label{eqn:A.4}
Z_{qq}^{\rm NS}(z) &=& \delta(1-z) + \frac{\alpha_s}{4\pi}
C_F \Big [-8(1-z)
+ \varepsilon \{ -8(1-z)\ln z 
\nonumber\\[2ex]
&& - 4(1-z)\ln(1-z) + 10 - 8 z) \} \Big ]
\nonumber\\[2ex]
&& + (\frac{\alpha_s}{4\pi})^2\Big [ C_F^2 \Big \{ 
-16 (1-z) - (8+16z) \ln z + 16(1-z)\ln^2 z
\nonumber\\[2ex]
&& - 16 (1 - z)\ln z\ln(1-z) 
\nonumber\\[2ex]
&& +C_AC_F \Big \{ -\frac{1}{\varepsilon}\frac{88}{3}(1-z)
-\frac{592}{9} (1-z) + 8(1-z)\,\zeta(2) 
\nonumber\\[2ex]
&& + (-\frac{80}{3}+\frac{8}{3}z)\ln z
- 4 (1-z)\ln^2 z \Big \}
\nonumber\\[2ex]
&&+n_fC_FT_f\Big \{ \frac{1}{\varepsilon}\frac{32}{3}(1-z)
+ \frac{16}{3}(1-z)\ln z + \frac{80}{9}(1-z)  \Big \} 
\nonumber\\[2ex]
&& - (C_F^2-\frac{1}{2}C_AC_F)\Big \{ 8(1+z)\Big ( 4 {\rm Li}_2(-z) 
+ 4 \ln z \ln(1+z) +2\zeta(2) 
\nonumber\\[2ex]
&& - \ln^2 z - 3\ln z \Big ) - 56(1 - z) \Big \}
\Big ] \,.
\end{eqnarray}
Where the last part originates from the process in Eq. (\ref{eqn:3.19}) with
two identical particles in the inclusive final state.
In the expressions above the definitions of the Riemann zeta-functions 
$\zeta(n)$ and the polylogarithms ${\rm Li}_n(z)$, ${\rm S}_{n,m}(z)$ can be 
found in \cite{lbmr}.

%


\end{document}